\documentclass[11pt]{article}
\usepackage{hyperref}
\usepackage{graphicx}
\pdfoutput=1
\begin{document}
\title{Elucidating the turbulence nature of the intracardiac flow: from medical images to multi-cycle Large Eddy Simulations}
\author{Christophe Chnafa $^1$, Simon Mendez $^2$, Franck Nicoud $^1$ \\
\\\vspace{6pt} $^1$ I3M, UMR CNRS 5149,  \\\vspace{6pt} $^2$CNRS and  I3M, UMR CNRS 5149,  \\ University Montpellier II, 34095 Montpellier, FRANCE}
\maketitle

\begin{abstract}
This brief article accompanies a \href{http://www.math.univ-montp2.fr/~yales2bio/IMAGES/APS_DFD2013/apsdfd102246.mp4}{fluid dynamics video}  presenting the results of a large-eddy simulation of the flow
in a realistic left heart. The left heart geometry, from the pulmonary veins to the aortic root, is extracted from medical images
and the endocardium movements are reconstructed through image registration. Large-eddy simulations are thus performed
in a patient-specific heart model, where the patient-specific movements of the geometry are prescribed. The flow obtained 
is intermittent, showing both in the left atrium and in the left ventricle turbulent spots correlated to flow decelerations. 
\end{abstract}

	\begin{figure}
	\centering
	\includegraphics[width=5.0in]{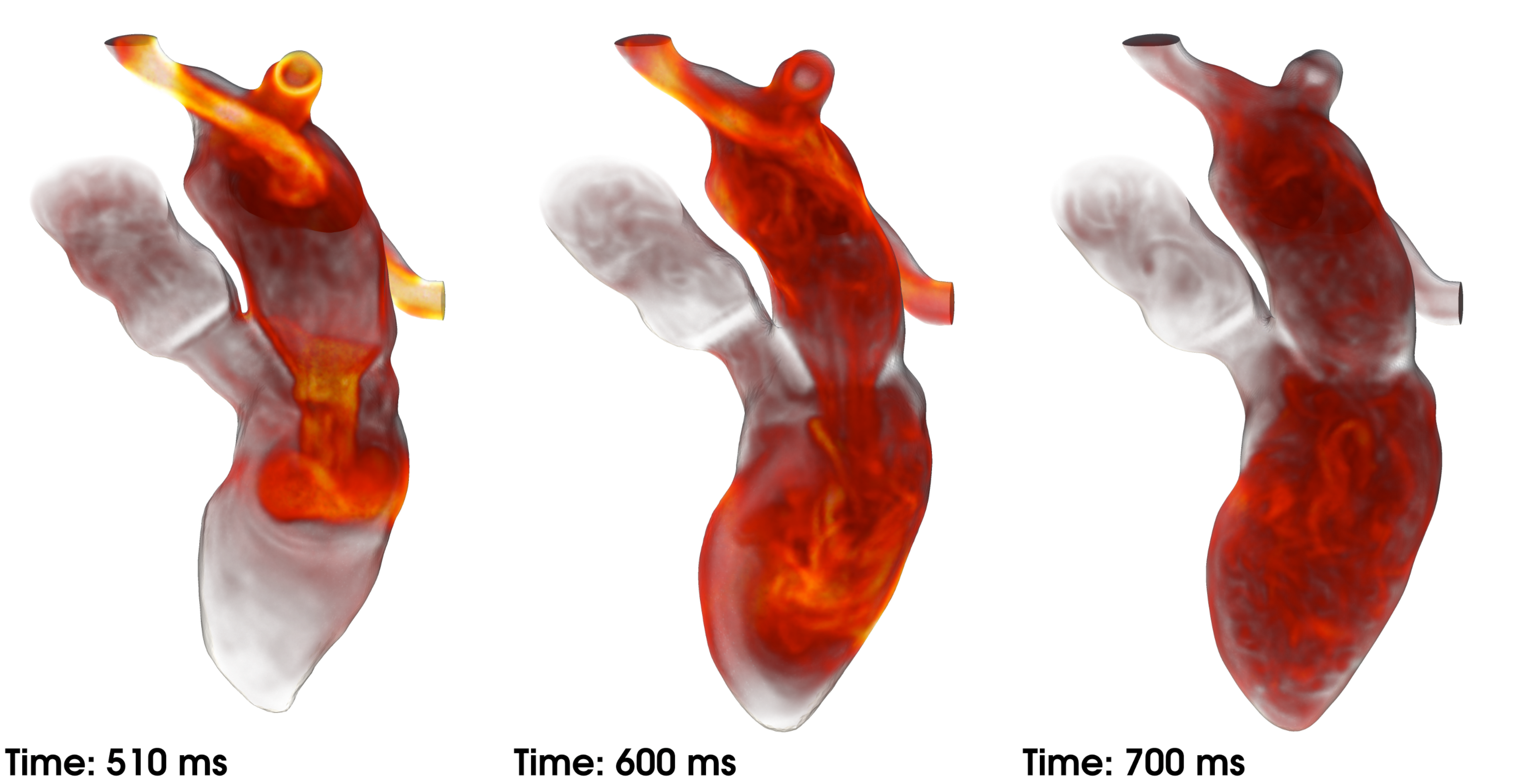}
	\caption{Vorticity magnitude in the left heart during diastole.}
	\label{fig:vorticity} 
	\end{figure}

\section{Introduction}

Although the blood flow in the left heart is inherently related to the heart function, this link is yet to be understood. 
The general organization of the flow can be found in textbooks \cite{Fung1997}, but the details of this three-dimensional
unsteady flow in a moving/deforming domain are still subjected to intense research. 
The improvement of medical imaging and numerical simulation techniques has given the physicians and the researchers
new insights in the left heart flow over the last years. 
 As a complement to medical imaging techniques like phase-contrast magnetic resonance imaging (PC-MRI) \cite{Markl2011} or echocardiography,
 we developed a numerical technique to perform numerical simulations of the left heart flow using morphological images. 
 
 The numerical procedure is presented in \cite{Chnafa:2013,Chnafa:2012,Midulla2012}. The video briefly explains the different steps 
 of the method, which are summarized here:
 \begin{enumerate}
\item A medical exam consisting of 4-D morphological images of a patient-specific left heart is indispensable for our method. CT-scan and MRI exams have already been successfully used. 
\item The images are segmented and a model of the left heart is constructed at one instant of the heart cycle.  In the case considered, it contains the left atrium and the left ventricle, the end of the four pulmonary veins and the aortic root. 
\item This computational domain  is discretized into an unstructured grid.
\item By image registration, the unstructured grid is deformed so that the computational domain follows the medical images. Interpolation is used to reconstruct the geometry between instants where medical images are available. From the knowledge of the wall displacements, the unstructured grid is deformed along the cardiac cycle using a mesh deformation algorithm. A time-evolving computational patient-specific computational domain is thus available.
\item Because of limitations in space-time resolution of medical data, valves cannot be accounted for in the same way and they are thus modeled. From medical images, we extract the location of the valvular annuli and the main characteristics of the mitral valve (length of the leaflets; position of the center, small and large radii of the best ellipse modelling the cross section area; periods of time when the valve is closed/open).  A simple geometrical model fed by these data is then used to assess the position of the leaflets over time and their effect on the blood flow is then accounted for thanks to an immersed boundary method.
 \item Mass flow rates are imposed at the four pulmonary veins (top of figure~1). We assume that either the mitral valve or the aortic valve is closed along the cardiac cycle, so that blood always enters in a closed domain. Mass conservation arguments can then be used to obtain the incoming mass flow rate, evenly distributed between the pulmonary veins. 
 \item Blood is modeled as a Newtonian fluid and Navier-Stokes equations are solved in a moving domain with prescribed boundary motion, using an Arbitrary Lagrangian-Eulerian framework. 
\end{enumerate}
 
 In the present \href{http://www.math.univ-montp2.fr/~yales2bio/IMAGES/APS_DFD2013/apsdfd102246_lowformat.mp4}{video} (see also the \href{http://www.math.univ-montp2.fr/~yales2bio/IMAGES/APS_DFD2013/apsdfd102246.mp4}{video in HD}), a CT-scan exam is used. The computational grid consists of 3 million tetrahedral cells, with a typical grid size of $0.8$~mm.  The maximum Reynolds number at the mitral valve during diastole and at the aortic valve during systole is of order of 5000. The heart rate is 60 beats per minute. Twenty-five heart cycles are computed and phase averages are gathered over 15 cycles. 
 The YALES2BIO solver \cite{Mendez2014} 
(\href{http://www.math.univ-montp2.fr/~yales2bio/}{http://www.math.univ-montp2.fr/$\sim$yales2bio/}) is used to perform the present large-eddy simulations. 
The non-dissipative 4th-order finite-volume numerical method, inherited from the YALES2 solver (\href{http://www.coria-cfd.fr/index.php/YALES2}{http://www.coria-cfd.fr/index.php/YALES2}) \cite{Malandin2013,Moureau2011a}, enables to detect turbulent spots along the cardiac cycle.

The simulation recovers the main characteristics of the left heart flow \cite{Chnafa:2013,Chnafa:2012}, as reported for instance in medical exams \cite{Markl2011} or \textit{in vitro} experiments.  
However, the simulation shows that, in addition of the usual picture of the left heart flow, turbulence may be present at specific regions and specific periods along the heart cycle. 
 Two regions in the left heart mainly show turbulence activity:  the upper part of the atrium and the lower part of the ventricle, near the apex. In both places, the flow destabilizes when the jets filling the cavities decelerate. In the ventricle, this occurs in the second half of diastole, after the impact of the vortex ring generated during the E-wave. Note that the A-wave is rather weak in this heart, which most probably  impacts the development of turbulence in the ventricle. 
 
An effective way of visualizing the flow features in the left heart is the volume rendering of vorticity magnitude, as shown in figure 1 and in the fluid dynamics videos.

\bibliographystyle{plain}
\bibliography{biblio}

\end{document}